# A Galactic Cosmic Ray Electron Intensity Increase of a factor of up to 100 At Energies between 3 and 50 MeV in the Heliosheath between the Termination Shock and the Heliopause Due to Solar Modulation As Measured by Voyager 1

W.R. Webber[1], N. Lal[2] and B. Heikkila[2]

1. New Mexico State University, Astronomy Department, Las Cruces, NM  88003, USA
2. NASA/Goddard Space Flight Center, Greenbelt, MD  20771, USA

**ABSTRACT**

We have derived background corrected intensities of 3-50 MeV galactic electrons observed by Voyager 1 as it passes through the heliosheath from 95 to 122 AU. The overall intensity change of the background corrected data from the inner to the outer boundary of the heliosheath is a maximum of a factor ~100 at 15 MeV. At lower energies this fractional change becomes less and the corrected electron spectra in the heliosheath becomes progressively steeper, reaching values ~ -2.5 for the spectral index just outside of the termination shock. At higher energies the spectra of electrons has an exponent changing from the negative LIS spectral index of -1.3 to values approaching zero in the heliosheath as a result of the solar modulation of the galactic electron component. The large modulation effects observed below ~100 MV are possible evidence for enhanced diffusion as part of the modulation process for electrons in the heliosheath.

**Introduction**

Galactic cosmic ray electrons have always been an enigma.  The recent studies of the LIS electron spectrum between ~3-70 MeV by Voyager (Cummings, et al., 2016) have continued this enigma in the sense that:  (1) How is it possible for this spectrum to be ~$E^{-1.3}$ as measured at Voyager, when at energies of a few GeV and above this electron spectrum is ~$E^{-3.2}$?  And also; (2) Why are the PAMELA electron intensity measurements at the Earth (Adriani, et al., 2015) at ~80 MeV only about 1/1000 of those measured at comparable energies at Voyager?

Regarding the $E^{-1.3}$ spectrum measured by Voyager, (Webber and Villa, 2017) have proposed that the spectral change from a LIS spectrum with index ~-3.2 above a few GeV to the one measured by Voyager at low energies can be explained by interstellar propagation effects.  Their argument is, that starting with an electron source spectrum, $dj/dE$ ~$E^{-2.25}$ below 5-10 GeV, this spectrum is modified by synchrotron and inverse Compton energy loss above ~1 GeV and also diffusion where $K$ ~$P^{0.5}$ and then by diffusion with a rigidity dependence ~$P^{-1.0}$ below a rigidity, $P_0$, which is equal to about 0.5 GV, at rigidities less than 1 GV.  The resulting "observed" LIS spectral index thus decreases as a result of this process from -3.2 at 5-10 GV rigidities (rigidity ≡ energy for electrons) to about -2.2 at 1 GV and then plunges to ~-1.3 below $P_0$.  The value of -2.25 for the source spectral index for electrons stays almost constant above and below 100 MeV and, in fact, is related to the choice of rigidity dependence of the diffusion coefficient below $P_0$.

The other enigma, the extreme amount of solar modulation of possibly as much as 1000 between Voyager and PAMELA electron measurements at ~60-80 MeV, requires an understanding of the modulation in the heliosheath.  The TET telescope and the HET-BSe mode on V1 (Stone, et al., 1977, 2013) have indeed measured large intensity increases between the termination shock and the heliopause (HP) in the heliosheath in the energy channels between 3.5 and 38.7 MeV.  However, because of relatively large non-electron "backgrounds" in these channels it has not been possible previously to obtain accurate flux values.

This paper attempts to evaluate these backgrounds and to determine flux values and electron spectra at increasing distances from the Sun as V1 moves from the HTS crossing at 95.0 AU to the HP crossing at 121.7 AU between 2005.0 and 2012.7 (Stone, et al., 2013).  The abruptness and decisiveness of this HP crossing in terms of the electron radial intensity gradients is also discussed.

## The Data:  The Electron Radial Intensity Profiles and the Background Corrections

The uncorrected intensity-time profiles of the 4 energy channels between 8.3 and 39 MeV on the TET telescope beyond about 90 AU are shown in Figure 1A.  This telescope is described in Stone, et al., 1977, and Cummings, et al., 2016, where spectra are presented.  The rate increases are large, ranging from a factor ~20 at the highest energy to more than 70 at the lowest energy.  However these rates cannot be used directly to obtain the absolute fluxes.  The background is as much as 50-80% of the total counts in each channel near the HTS.

The intensity-time profiles of HET-BSe in 3 energy channels between ~2.5 and 12 MeV are shown in Figure 1B.  This telescope is described in Cummings, et al., 2016, where spectra are presented.  Here the increases are also large, ranging from factors ~20-25 between the HTS crossing at 2005.0 and the HP crossing at 2012.65.  For these lower energy electrons the background estimation proceeds in a different way as described below.

### (A):  TET Background Calculations

Fortunately the TET telescope background can be estimated from a large solar flare electron event in April, 1978, when V1 was at ~2.5 AU and also using the GEANT4 calculations.  A matrix of events D1 vs. D2, which comprise the total uncorrected counting rates used in the intensity time profiles in Figure 1A is shown in Figure 2, not included because of size limitations.  It is obvious that most of the counted events are electrons distributed around a peak channel ~45 in each D counter.  If we overlay a grid on this matrix that extends along the D1 vs. D2 diagonal and extends in the perpendicular direction from this diagonal so as to include ~95% of the total electron events, which is the fraction of electrons based on the D1-D2 electron distribution that lies within this grid, we find that only 86% of all events on the matrix are electrons, that is they lie inside this grid.  The other 14% are events scattered throughout the D1-D2 matrix including events along the D1-D2 diagonal but with pulse heights >2 times minimum.  The distribution of events along the grid diagonal itself for each of the 4 energy intervals is shown in Figure 3.   These distributions are taken to be the "reference" pulse height distributions along the diagonal for a pure electron component.

This same grid is then applied to all selected 52 day average TET matrices for all energy intervals during the time period V1 passes through the heliosheath.  Examples of D1 vs. D2 matrices with the grid on them are shown in Figures 4A and B for the 8.3 MeV energy channel for the time periods centered at 2005.5, just after V1 entered the heliosheath, and at 2012.1, about 2 AU inside the HP, but are not included in this submission due to size limitations.  The distribution of events along the D1 and D2 diagonal as obtained from these matrices are shown in Figures 5A and B for the 8.3 MeV average energy (in red) and also for the higher energies, 15.4, 25.2 and 38.7 MeV, in black, blue and green respectively, for these same two time periods in Figures 5A and B respectively.  The electron "peak" that is prominent at 8.3 MeV becomes somewhat less prominent but is still evident at higher energies as shown in Figures 5A and B for the individual histograms for each energy and also by comparing these histograms at the 2 times, one just outside the HTS, and the other just inside the heliopause.  Near the HTS a weaker electron peak still exists even for the highest energies.  The electron fractions for each time interval and for each energy are determined from the counts in the shaded regions in Figures 5A and B, shown here for the 8.3 MeV energy interval.  The horizontal solid lines in the histograms represent the estimated background at each energy, assuming a constant number of events per pulse height interval underneath the peaks (most of the backgrounds are consistent with a constant background at pulse heights above ~1.8 times min).  Possible variations from this initial constant background estimation as obtained from the individual background distributions, including an increasing background underneath the peaks, are included in the error estimates.

The electron fractions and corrected intensities in each energy and time interval for the TET telescopes are shown in Table 1 for times centered at 2004.8, 2005.5, 2007.7, 2009.2, 2010.6 2012.1 and also for an interstellar time period as V1 moves from just inside the HTS to just beyond the HP.  These electron fractions range from ~50% just outside the HTS to 86% in the LIS medium for the lowest energy, to ~22% to 67%, respectively, for the highest energy channel.

Using these electron fractions and the total electron rates that were shown in Figure 1A we may derive the ratios of the intensities at each location to the LIS intensities reported by Cummings, et al., 2016.  These ratios, at each energy, are shown in Figure 6.  Note that at ~120 AU, just 1.7 AU inside the HP, the ratio to the LIS spectrum is an almost constant value ~0.5 at each energy, indicating a rigidity independent amount of solar modulation. In other words, an intensity change (radial gradient) ~50% per AU is observed at all energies from 8.3 MeV to 39 MeV at this location near the heliopause. As one goes deeper in the heliosheath, from the HP inward to the HTS, a minimum develops in the ratio. The 15.4 MeV value becomes the lowest ratio, even lower than the ratio at 8.3 MeV. A relative "excess" thus appears in the lowest energy channel.

### (B) Estimation of the HET-BS Electron Backgrounds

The background correction for HET-BSe is essentially a two-step process. We first make a HET, B1 vs. B2 matrix of events, in which the electrons have energy loss $\equiv$ pulse height equivalent to that of a relativistic particle with an average pulse height channel ~4, which lies in a corner of the matrix which is set up for stopping proton and helium energy loss. A box is defined with criteria 3<(B1+B2)<16 and (-6.0+0.92*B2)$\leq$B1<(4.5+0.92*B2). This "separates" electrons from other events on this matrix. This rate of events has a "background" which is small at the high rates present in the LIM but becomes larger as one goes deeper into the heliosheath and the rates decrease.

The electrons then penetrate the C432 stack of counters producing a distribution of events that satisfy the B1-B2 criteria and produce pulse heights in C432 that are related to their energy, e.g., channel 1 = 3.5 MeV, channel 8 = 10 MeV, on average. There is a background at larger C4 channels in the C4 vs. (B1+B2) matrix. In fact, the distribution of events in the C4 channel in this matrix is shown in Figure 7 for the sum of the time intervals 2004.8 and 2005.6 where the electron intensity in the heliosheath is at a minimum. This background can be seen at higher channels with a much flatter slope of the distribution of pulse heights. This background amounts to as much as 20-30% at channels corresponding to electron energies ~10 MeV which is the case for time intervals where the electron intensity is smallest near the HTS. The background corrected fractions for HET-BSe relative to the LIS are shown as open circles on Figure 6. Note that the corrected HET-BSe fractions match well with those for TET at a corresponding energy at 10 MeV.

**A Discussion of the Modulated Electron Spectrum in the Heliosheath**

To interpret the ratios shown in Figure 6 from the TET and HET BSe telescope in more detail, we use the ratios in Figure 6 and the LIS intensities derived by Cummings, et al., 2016, and modified for the new LIS background fractions derived in this paper (see bottom of Table 1) to derive the "relative" intensities of electrons at all energies at several radial locations in the heliosheath.

These intensities are shown in Table 1 and in Figure 8A in a j x $E^2$ format and in Figure 8B in a normal intensity vs. energy format, along with the PAMELA measurements of electrons at the Earth down to ~80 MeV. A systematic decrease in intensity with decreasing radius along with an increasingly flatter spectrum (in the sense of a smaller negative spectral exponent) is noted above ~8 MeV in Figure 8A in which the intensities are times $E^2$. Near the HTS at 95.0 AU (the shaded regions in Figure 8A and B) the galactic electron spectrum between 15 and 40 MeV has an "average" slope of -0.3, which is about 1.0 power less in negative slope than the LIS electron spectrum which is ~$E^{-1.3}$. This difference in slope is due to solar modulation in the heliosheath. Note that the slope of the average PAMELA electron spectrum on the same figure between ~120-300 MeV is about +0.5. These differences between these spectral exponents near the HTS and the Earth represent features of the solar modulation inside the HP.

At energies below ~10 MeV there is a distinct change in the electron spectra. These differences, best seen in Figure 8A, lead to spectra with increasing negative spectral index with increasing distance inward from the HP reaching an index ~-2.0 at 114.5 AU and becoming still steeper, with an index ~-2.5 near the HTS.

**Radial Gradients in the Heliosheath – A Discontinuity at the Heliopause**

It is seen in both Figures 6 and 8A and 8B that at ~15 MeV the decrease in intensity from the HP inward to the HTS is a factor ~100. Over a radial distance ~27 AU, which is the thickness of the HS, this corresponds to an average radial gradient ~20%/AU. In other words the intensity doubles every 4 AU with a final doubling of intensity over the final 1.7 AU just before the HP.

This final increase between 2012.1 and the LIS intensity after 2012.65 at 8.3 MeV is shown in Figure 9A. This final factor of ~2 increase after 2012.1, was seen at all energies as noted earlier. Figure 9B is an expanded version of 9A in the vertical direction showing intensity changes as small as ± 1%. The statistical uncertainty of each 52 day average being used is ± 0.5%. The radial intensity gradient at 8.3 MeV changes in one 52 day interval from ~88%/AU between the last two 52 day time intervals separated by 0.14 AU, to ~0.1%/AU or less for the time period beyond ~2012.7. Figure 9C presents the same scaling as 9B, but for the 2.5-5.2 MeV BS electrons. Here the radial gradient changes from ~130%/AU to ~0.1%/AU or less in about one 26 day time interval. This very distinct boundary between heliospheric modulation and interstellar features is recognizable as the heliopause.

## A Comparison of Electron and Proton Modulation at V1 in the Heliosheath

In a recent paper using both V1 and V2 data (Webber, et al., 2017) we have studied the proton intensities and solar modulation throughout the heliosphere including the heliosheath. In Table 2 and Figure 10 of this paper we compare the proton intensities at 250 MeV reported in the above paper, and the electron intensities at 3.5, 8.3 and 39 MeV that are measured at V1 in this paper during the 7 time intervals beyond the HTS. The intensities of the two species of particles, differing in rigidity by a factor ~10-100, are nevertheless well correlated at all energies and the regression lines at each energy for electrons are similar and indicate fractional changes in electron intensities that are approximately 18.5 times larger at 3.5 MeV, 25 times larger at 8.3 MeV and 15.5 times larger at 39 MeV than those of the protons of 250 MeV at a higher rigidity = 720 MV.

## Summary and Conclusions

In this paper we evaluate the backgrounds in the TET and HET-BSe electron telescopes on V1 in order to determine the true electron intensities measured at V1 as it passes through the heliosheath at distances between 95 and 122 AU from the Sun. Four energy ranges from 8.3 to 39 MeV are used in this analysis for the TET telescope along with four energies from 3.5 to 9.9 MeV for the HET-BSe telescope. This study is greatly facilitated by analyzing the response of these telescopes to a large solar flare electron event that occurred in April 1978, just 8 months after the launch of V1.

The electron fractions of the total TET telescope rates shown in Table 1 of this paper range from ~50% just outside the HTS to about 88% for the LIS spectrum for the lowest energy channel at 8.3 MeV to electron fractions of 22% and 67%, respectively, for the highest energy channel at 39 MeV.

These fractions are intensity dependent as would be expected since the electron rates are changing much more rapidly than the protons of a few 100 MeV that are producing much of the background.

For the HET-BSe telescope, the backgrounds are determined from the E-loss distributions of these events in the C432 counter (see Figure 7) and are energy dependent ranging from ≤ 5% at the lower energies of ~3.5 MeV to as much as 30% at 10 MeV.  These background corrected fractions relative to the LIS intensities for the TET and HET-BSe telescopes are shown in Figure 6.

The corrected electron intensities in each 52 day interval shown in Figures 8A and 8B are found to increase by over a factor of 100 from just outside the HTS to the LIS at an energy of 15.4 MeV.  There is an apparent "turn up" in the electron energy spectra at energies below ~10 MeV.  This turn up extends to the HET-BSe spectra at lower energies.  These heliosheath spectra steepen from the LIS spectra which has a spectral index equal to -1.3 to spectra with indices -2.0-2.5 at locations just beyond the HTS.

The galactic component of the electron spectrum in the heliosheath above 10 MeV becomes much flatter than the LIS spectrum as one goes further inwards and may begin to show features of the variation of the local diffusion coefficient with rigidity at rigidities below 100 MV, a feature to which the modulated electron spectrum is very sensitive.

The modulated galactic electron spectrum between 15 and 40 MeV near the HTS at ~95 AU is still, however, about a factor ~10 above the PAMELA electron spectrum at the Earth projected below ~100 MV as can be seen from Figure 7A.  From these values it appears that ~90% or more of the total electron modulation of a factor of 1000 in the heliosphere below ~100 MV observed by PAMELA at the Earth, actually occurs in the heliosheath.

The solar modulation effects on low energy electrons and much higher energy protons are well correlated in the heliosheath; however, the changes in electron intensities are much larger than those of the higher rigidity protons. These electron intensity <u>changes</u> are factors of 18.5, 25.0 and 15.6 times greater at 3.5, 8.3 and 38.7 MeV, respectively, than the proton changes at 250 MeV (720 MV). These relative changes as a function of rigidity pose problems for modulation theories and the rigidity dependence of the diffusion coefficient at these lower rigidities. Also the origin of the electrons which comprise the turn up in the spectrum, below ~10 MeV is uncertain. What fraction of these electrons are modulated galactic electrons and what part is heliosheath accelerated electrons?

**<u>Acknowledgments:</u>** The authors are grateful to the Voyager team that designed and built the CRS experiment with the hope that one day it would measure the galactic spectra of nuclei and electrons. This includes the present team with Ed Stone as PI, Alan Cummings, Nand Lal and Bryant Heikkila, and to others who are no longer members of the team, F.B. McDonald and R.E. Vogt. Their prescience will not be forgotten. This work has been supported throughout the more than 35 years since the launch of Voyager by JPL.

| TIME | R (AU) | | ENERGY INTERVAL | | | |
|------|--------|--|-----------------|--|--|--|
| | | | **8.3 MeV** | **15.4 MeV** | **25.6 MeV** | **38.7 MeV** |
| 2004.8 | 94.5 | FRACTIONS (%) | 50.2 (4) | 34.6 (5) | 31.5 (5) | 24.2 (5) |
| | | INTENSITY | 3.87 | 0.865 | 0.652 | 0.70 |
| 2005.5 | 97 | FRACTIONS (%) | 50.3 (4) | 34.0 (4) | 28.7 (5) | 20.2 (5) |
| | | INTENSITY | 4.50 | 1.06 | 0.815 | 0.718 |
| 2007.65 | 104 | FRACTIONS (%) | 67.2 (3) | 49.1 (3) | 35.0 (4) | 30.3 (4) |
| | | INTENSITY | 10.56 | 2.97 | 1.93 | 1.66 |
| 2009.15 | 110 | FRACTIONS (%) | 77.0 (3) | 65.5 (3) | 55.8 (4) | 47.2 (4) |
| | | INTENSITY | 37.85 | 11.95 | 6.56 | 4.8 |
| 2010.5 | 114.5 | FRACTIONS (%) | 74.3 (3) | 72.7 (3) | 67.5 (4) | 54.6 (4) |
| | | INTENSITY | 74.8 | 29.8 | 16.8 | 9.8 |
| 2012.07 | 120 | FRACTIONS (%) | 81.0 (3) | 79.0 (3) | 73.4 (4) | 64.4 (4) |
| | | INTENSITY | 152.2 | 62.6 | 32.8 | 19.4 |
| 2014.1 | 126.5 | FRACTIONS (%) | 86.1 (3) | 82.3 (3) | 74.5 (4) | 67.8 (4) |
| | | INTENSITY | 326 | 131 | 68.1 | 35.8 |

**TABLE I
ELECTRON FRACTIONS (%)**

These LIS are for 52 day intervals centered on the time indicated. The values in parenthesis are the errors in the percentage values, e.g., at 8.3 MeV the fraction is $50.2 \pm 4.0$. The intensities are in electrons/$m^2 \cdot sr \cdot s \cdot MeV$.

| Background Fraction In the LIS interval, 2014.1 | -13.9% | -17.7% | -25.5% | -32.2% |
|---|---|---|---|---|
| Background Assumed In Cummings, et al., 2016 | -2.1% | -4.1% | -8.6% | -16.9% |
| New Intensity Fraction in LIS | 0.879 x ApJ | 0.824 x ApJ | 0.815 x ApJ | 0.816 x ApJ |

| TABLE II | | | | | | | | |
|---|---|---|---|---|---|---|---|---|
| **INTERVAL** | **H** | **φ** | **e** | | | **RATIO** | | |
| | **250 MeV J** | **MV** | **3.5 MeV j** | **8.3 MeV j** | **38.7 MeV j** | **3.5e 250 H** | **8.3e 250 H** | **38.7e 250 H** |
| 2004.8 | 2.22 | 310 | 52 | 3.73 | 0.84 | 23.4 | 1.68 | 0.378 |
| 2005.5 | 3.56 | 210 | 65 | 4.46 | 0.82 | 18.3 | 1.25 | 0.230 |
| 2007.65 | 5.35 | 135 | 136 | 11.56 | 2.11 | 25.4 | 2.16 | 0.394 |
| 2009.15 | 6.57 | 100 | 296 | 39.3 | 5.50 | 48.1 | 5.98 | 0.837 |
| 2010.5 | 7.55 | 80 | 474 | 78 | 10.5 | 62.8 | 10.3 | 1.39 |
| 2012.07 | 8.81 | 53 | 610 | 166 | 21.3 | 69.2 | 18.8 | 2.42 |
| LIS | 13.1 | 0 | 1050 | 326 | 41.4 | 80 | 24.9 | 3.16 |

j in units of particles/m$^2 \cdot$ sr$\cdot$ s$\cdot$ MeV

**FIGURE CAPTIONS**

**Figure 1A:** Uncorrected TET intensity-time profiles of cosmic ray electrons in the heliosheath at V1. The mean energies of the rates are: (1)=8.3 MeV; (2)=15.4 MeV; (3)=25.6 MeV; and (4)=39.5 MeV.

**Figure 1B:** The same as Figure 1A except for lower energy electrons at: BS4e=3.8 MeV, BS3e=6.6 MeV and BS2e=9.9 MeV.

**Figure 2:** A matrix of D1 vs. D2 events for the 8.3 MeV channel for the solar flare electron event of April 21-24, 1978. The distribution of events (energy loss) about channel 45 in each dimension is due to solar electrons. The diagonal "grid" used in this analysis is superimposed on this matrix.

**Figure 3:** Distributions of events along the diagonal in the solar flare electron event of April 21-24, 1978. The excess above the background for the 8.3 MeV interval is shaded in red above a background of 30 counts/bin, for the 6 grid intervals between 1 and 2 times minimum energy where all of the electrons are observed. The distribution for the 15.4 MeV interval is in black and would be shaded above ~24 counts/bin, for the 25.6 MeV interval in blue and the shading would also be above ~24 counts/bin, for the 39.5 MeV interval the shading would be above a background of ~20 counts/bin

**Figures 4A and 4B:** Examples of matrices with the diagonal grid on them for the 52 day time periods centered at 2005.5 and 2012.1, when V1 is just outside the HTS and just inside the HP are shown. The remaining events on the outside of the diagonal grid are background.

**Figures 5A and 5B:** Distributions of events along the diagonal as determined for the two time intervals shown in Figures 4A and 4B. The true electrons are determined from the area under the distribution curves as shown, for example, by the shaded regions above the "diagonal" background for 8.3 MeV electrons which is assumed to be a constant = 190 counts/bin below ~2 times minimum in 5A and = 115 counts/bin in 5B. For 15.4 MeV electrons, these values are respectively 190 and 100, for 25.6 MeV electrons, 200 and 100, for 38.7 MeV electrons 240 and 100.

**Figure 6:** Ratios of the background corrected intensities to those determined for the LIS spectrum in Cummings, et al., 2016. Open circles are from HET-BSe, solid circles from TET. Data obtained just after launch in 1977 when Jovian electrons were present at lower energies is also shown in red.

**Figure 7:** Distribution of events in C432 channels for HET-BSe. Each channel has a width ~1.2 MeV so the 1$^{st}$ 10 channels represent electrons between about 2-12 MeV. The slope of the "background" determined at larger values of C4 and is extended to lower energies (dashed line). This background is mainly related to low energy protons also present on the (B1+B2) vs. C4 matrix.

**Figures 8A and 8B:** Electron intensities and spectra determined in the heliosheath from the ratios of intensities to the LIS intensities presented in Figure 6. Also shown is the PAMELA electron spectrum measured in late 2009 (Adriani, et al., 2015) at a time of extremely low solar modulation at the Earth. The intensities, j, are times E$^2$ in Figure 8A and Figure 8B is a normal j vs. E plot.

**Figures 9A, 9B and 9C:** Figure 9A shows the intensity of 8.3 MeV electrons for the time period from 2012.0 to 2017.0. The time periods are 52 days which have a statistical uncertainty of $\pm$ 0.5%. In Figure 9B the vertical scale is expanded so that variations ~ $\pm$ 1% are observable. Figure 9C shows similar data for 3.8 MeV electrons with an expanded vertical scale.

**Figure 10:** Regression curves comparing the 250 MeV proton intensities and those for 3.5, 8.3 and 39 MeV electrons beyond 95 AU in the heliosheath all measured at V1. The dashed lines approximating the 3.5, 8.3 and 39 MeV electron intensity increase in the HS have slopes that are 18.5, 25.0 and 15.5, respectively, times that of the proton intensity variations.

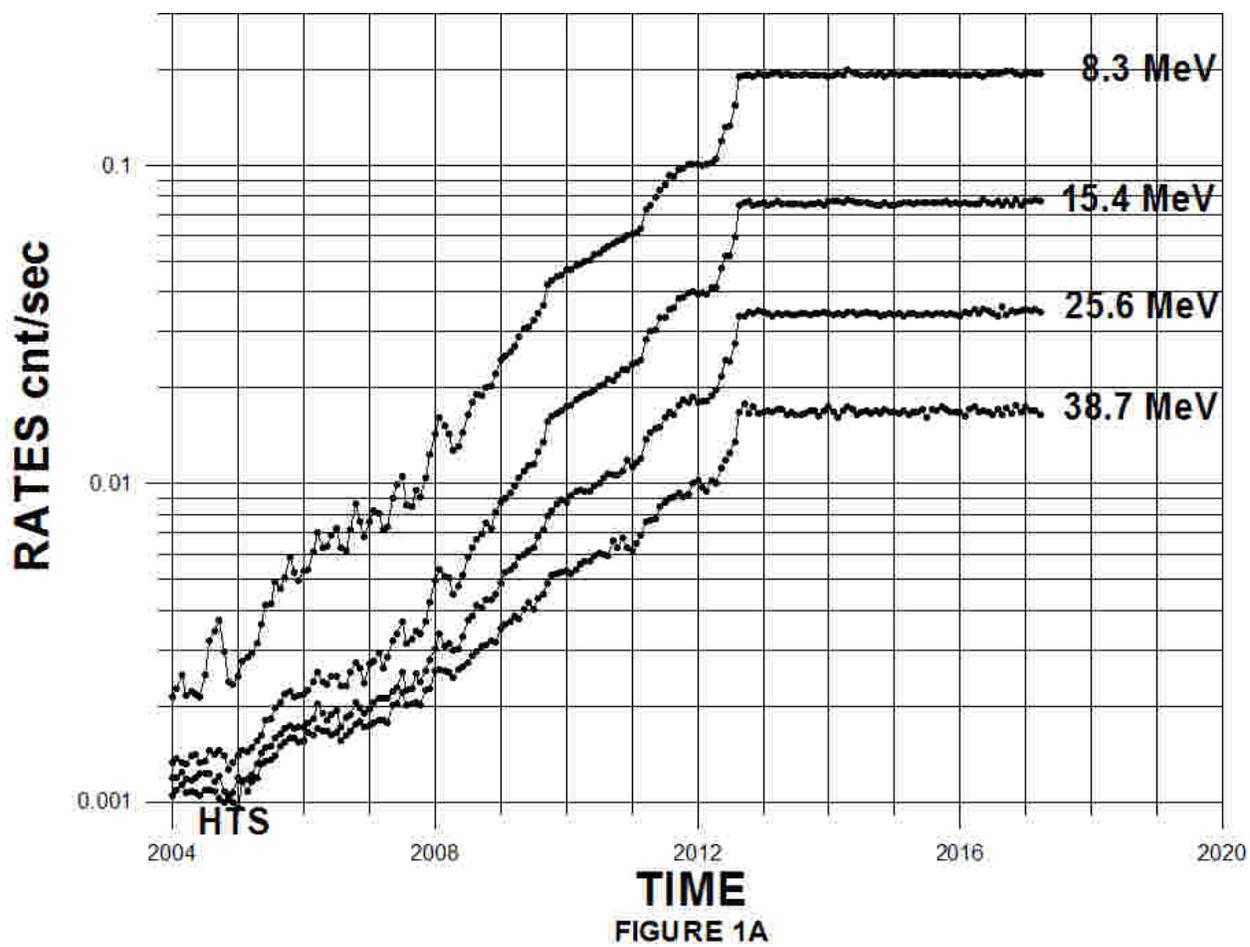

FIGURE 1A

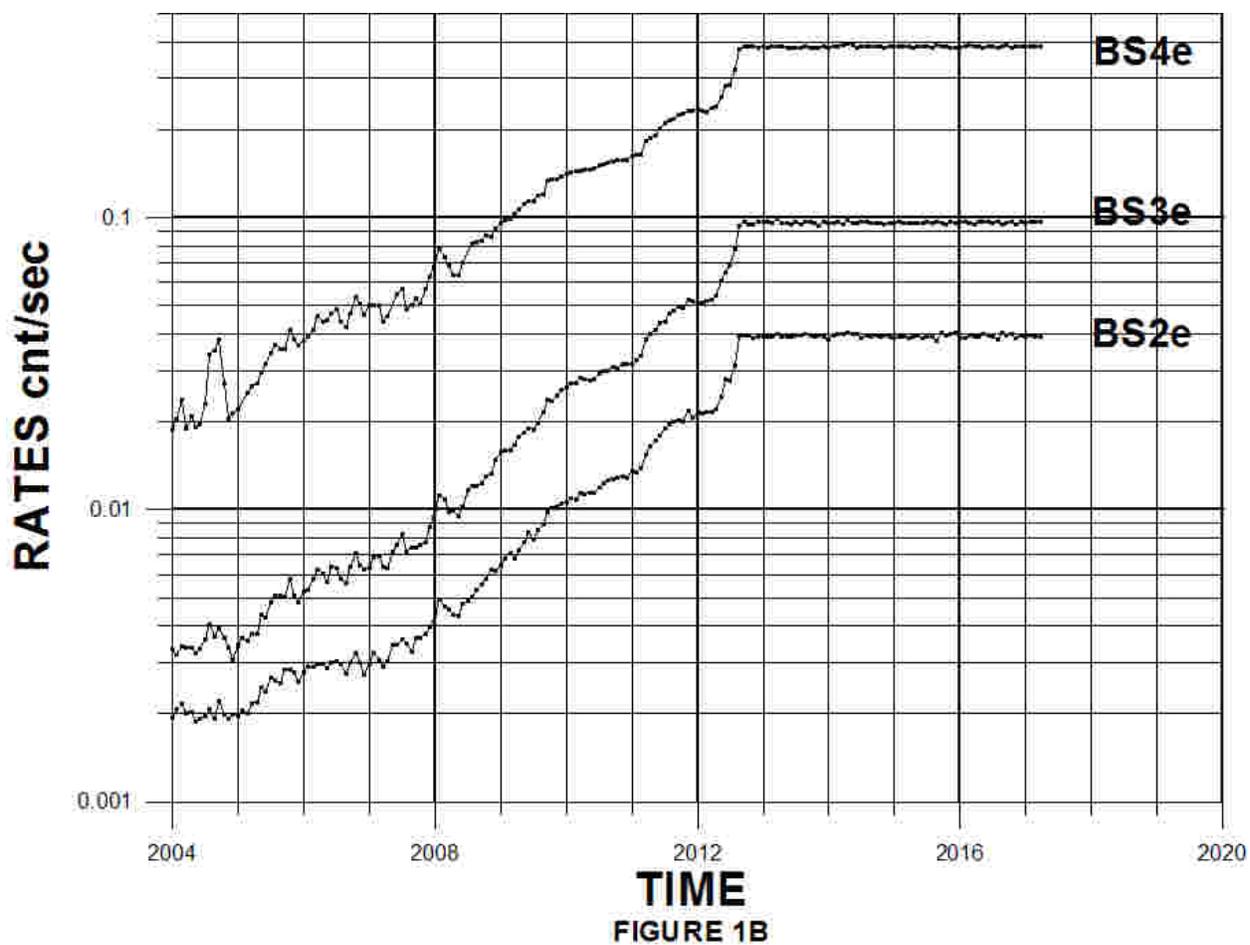

**TIME**

FIGURE 1B

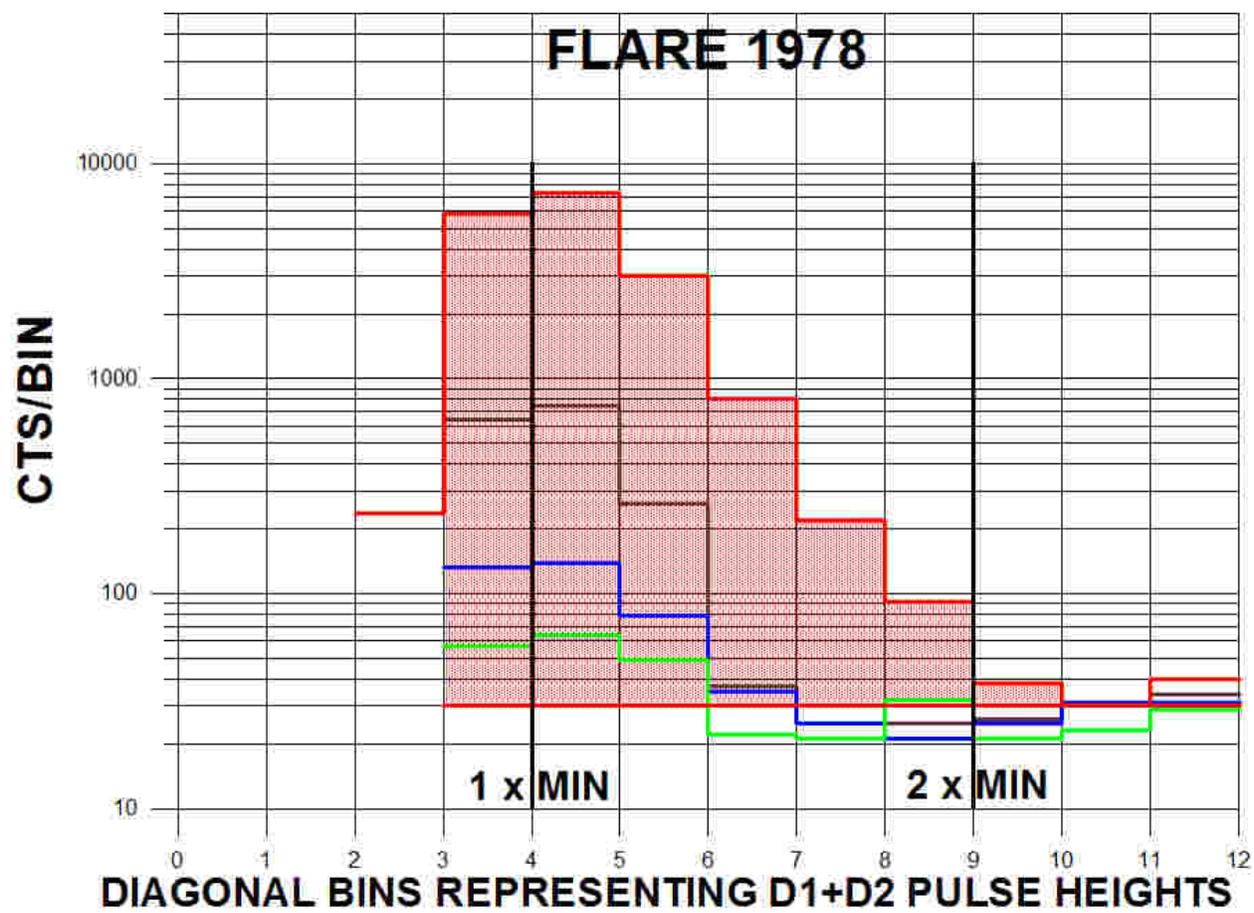

FIGURE 3

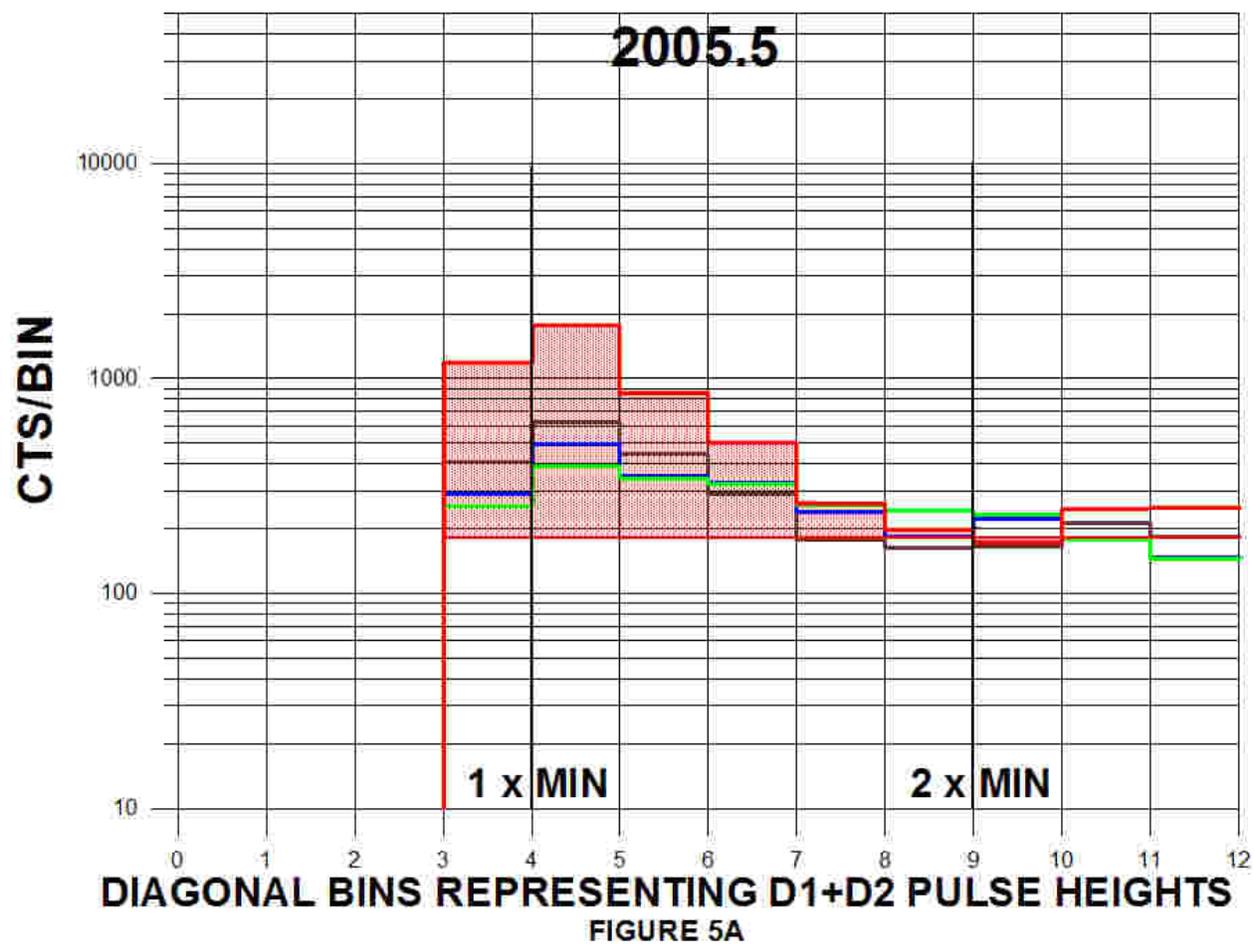

FIGURE 5A

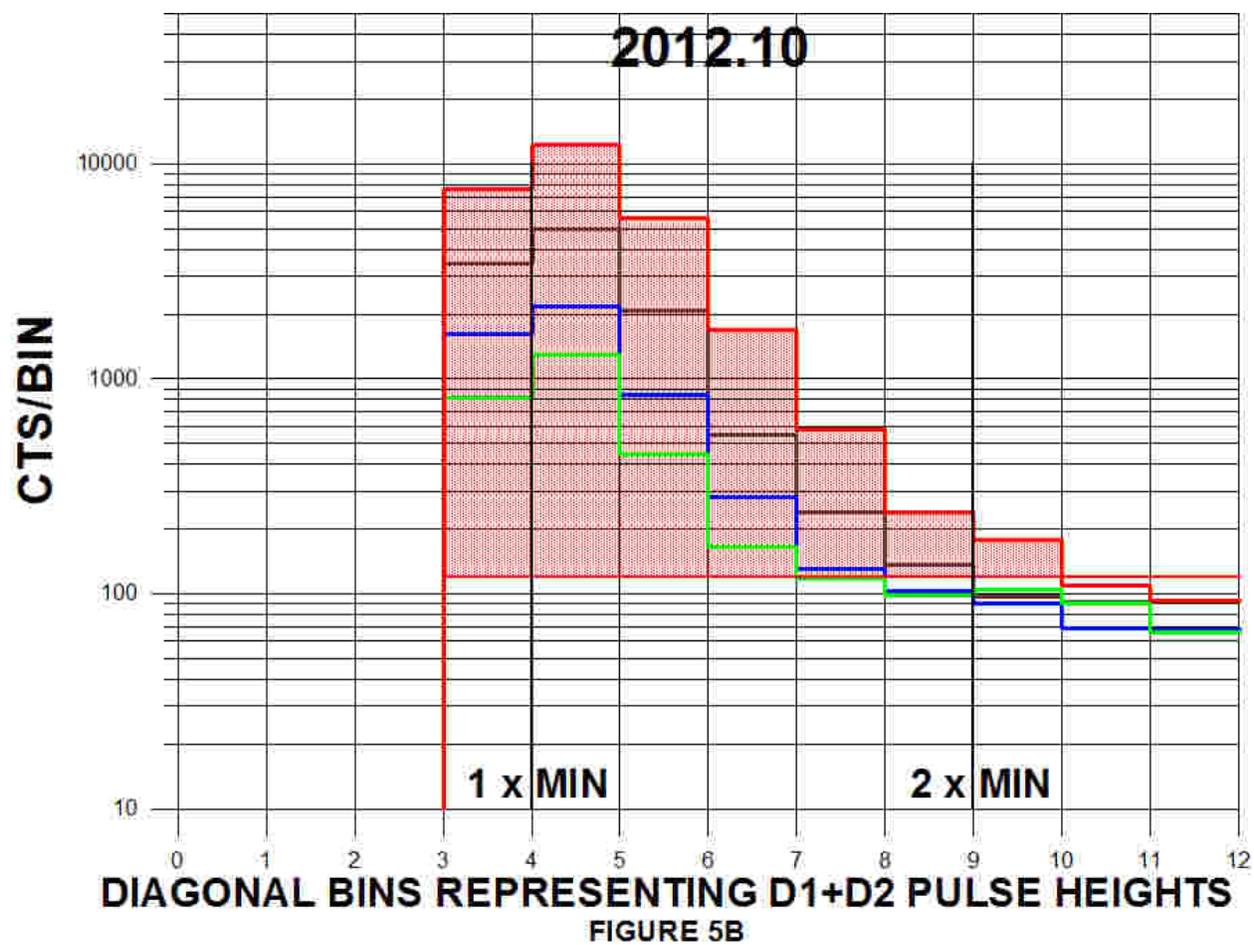

FIGURE 5B

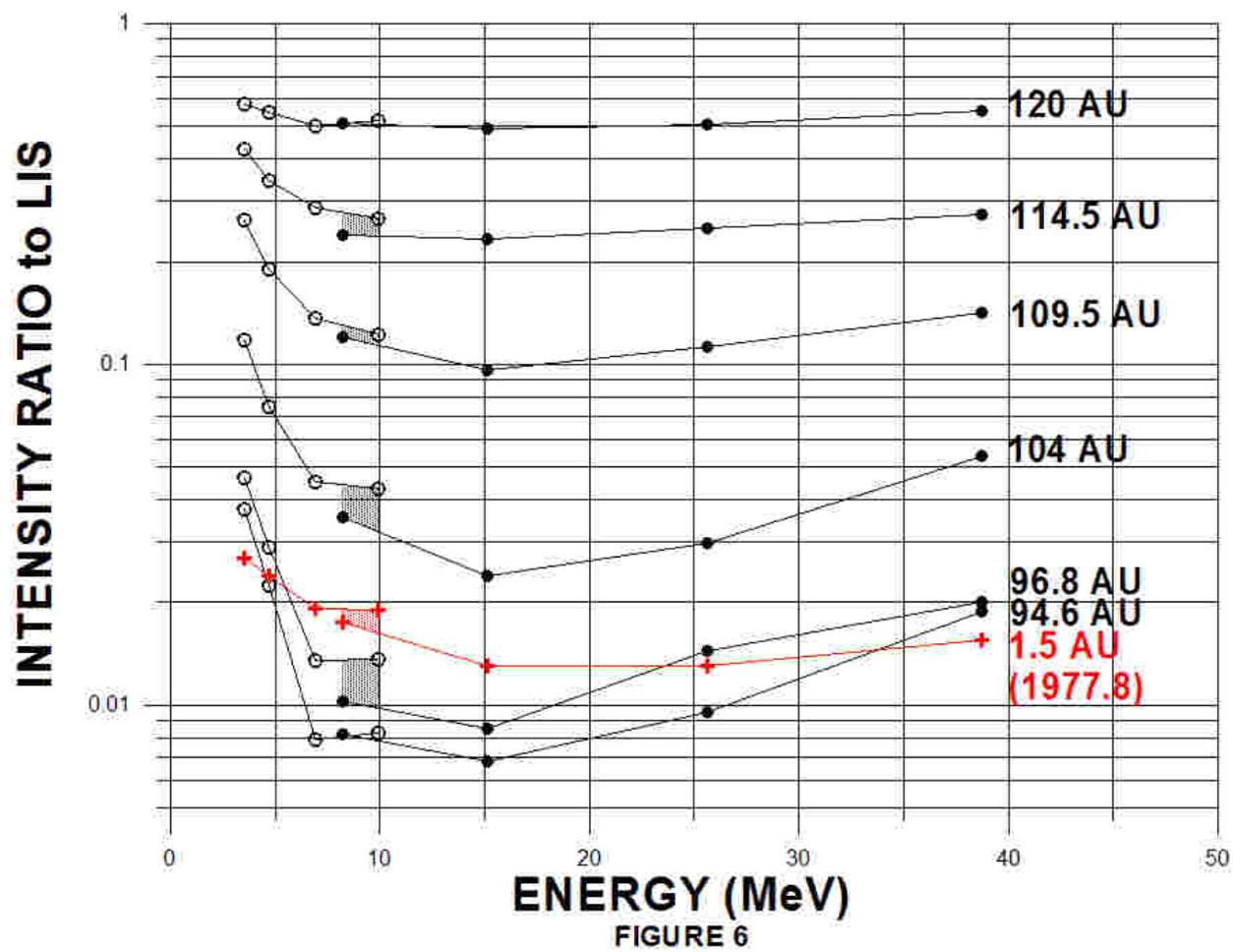

**FIGURE 6**

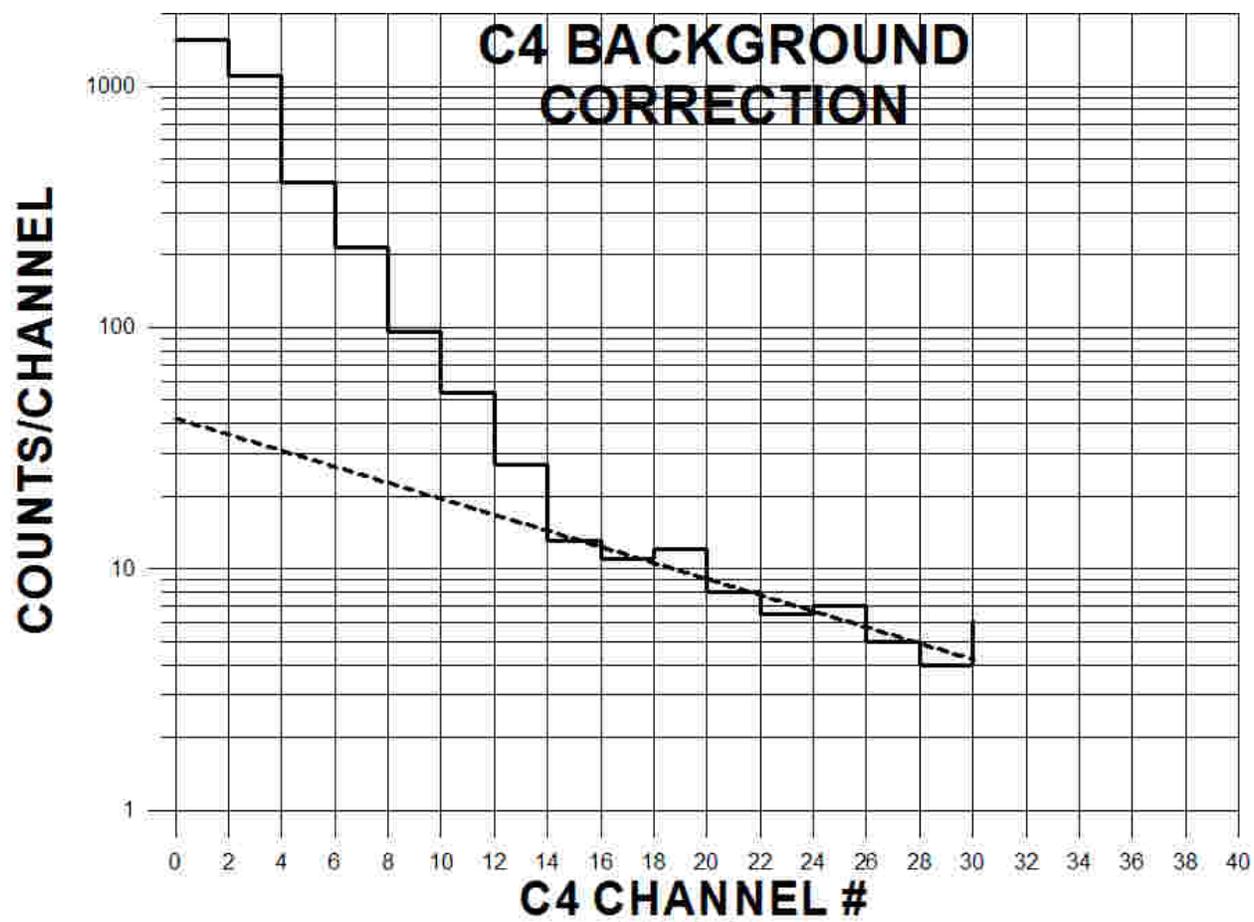

**C4 BACKGROUND CORRECTION**

FIGURE 7

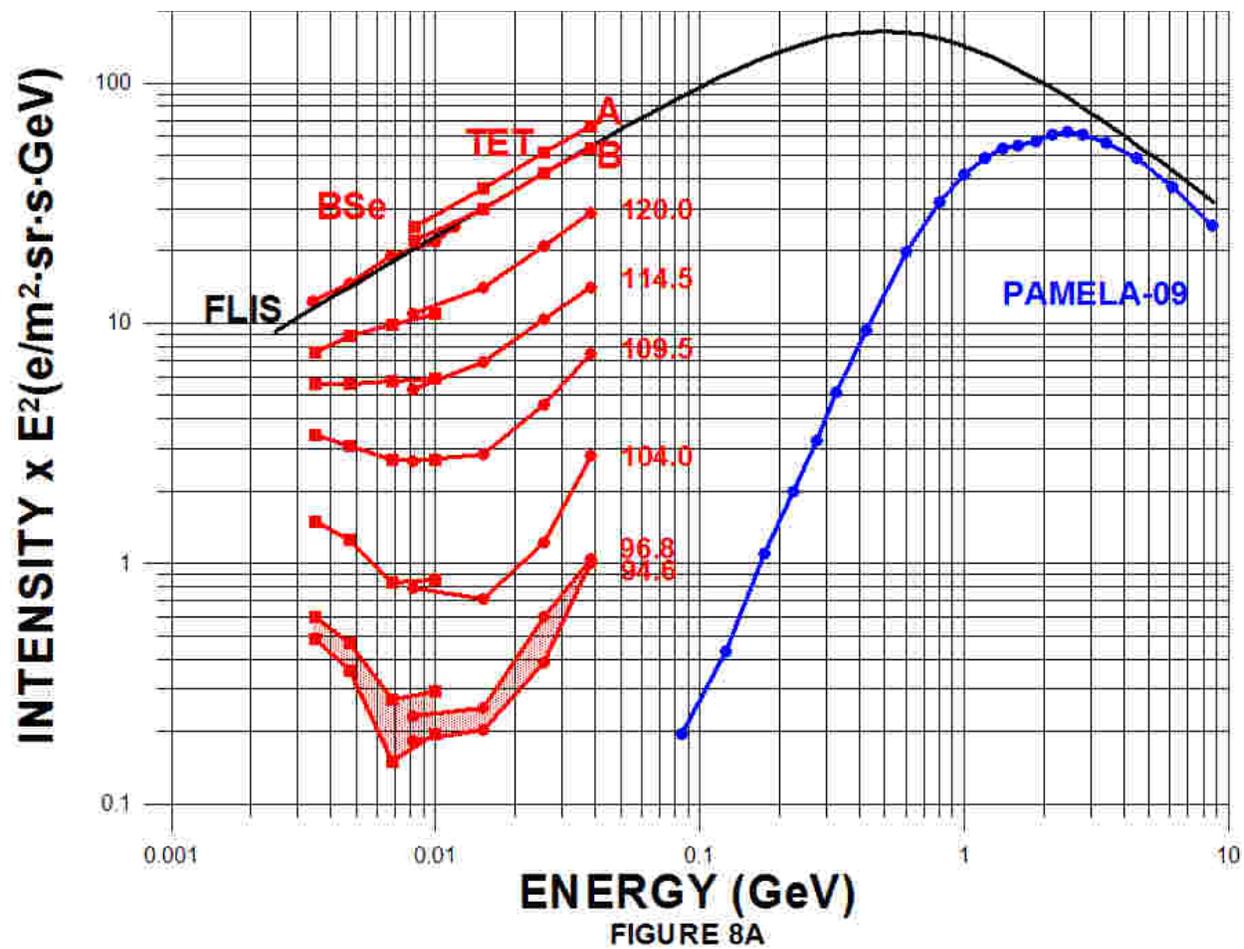

FIGURE 8A

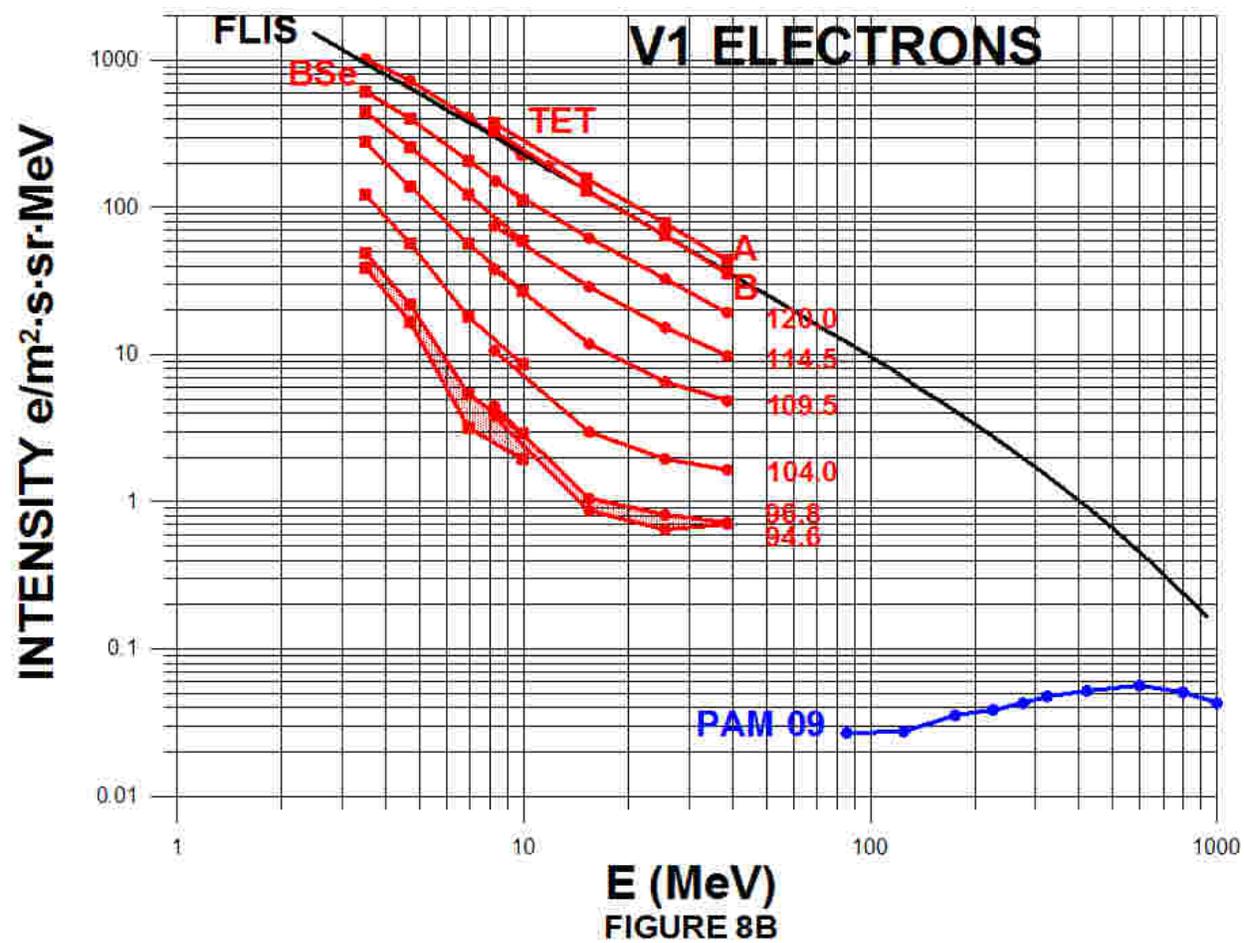

FIGURE 8B

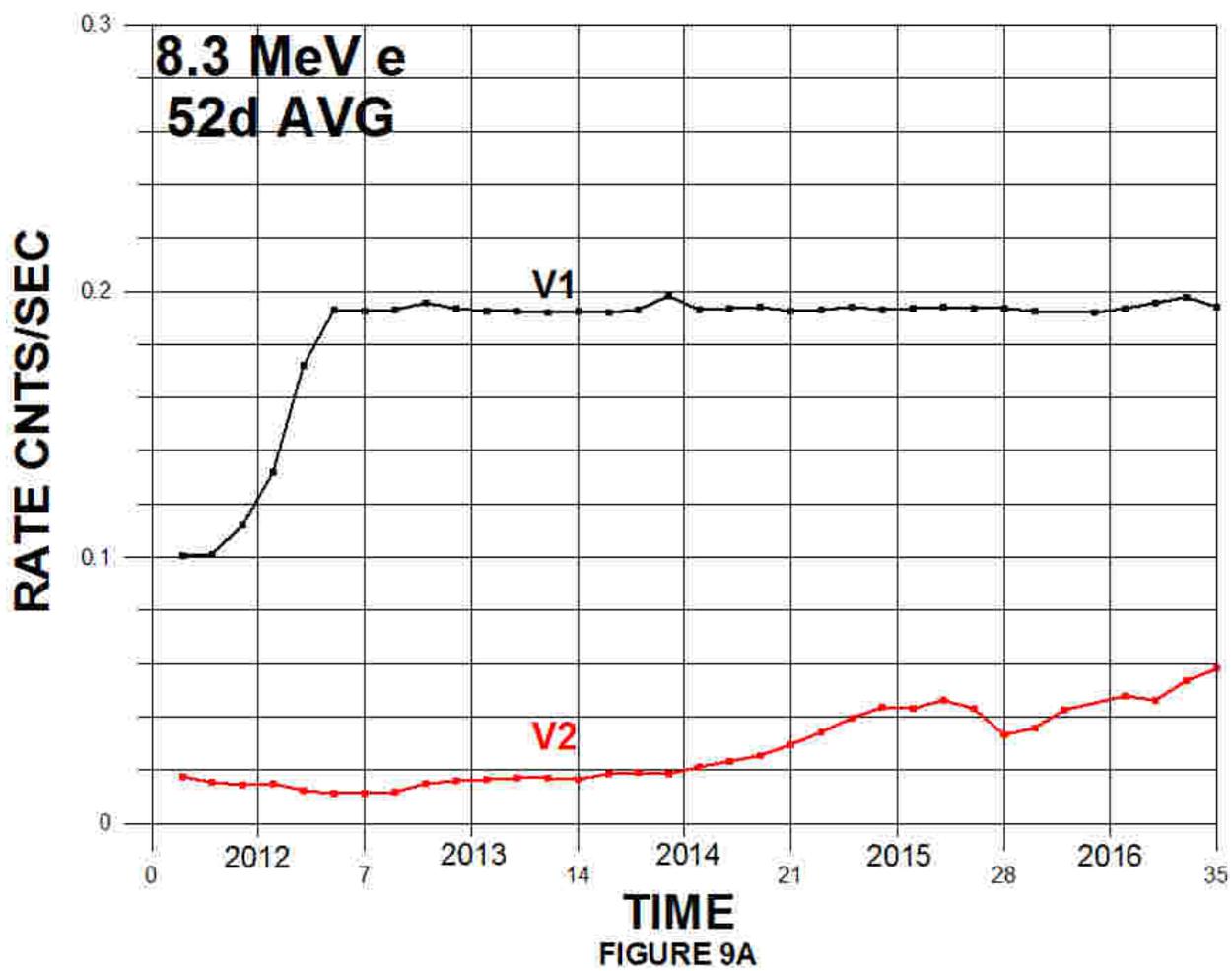

**FIGURE 9A**

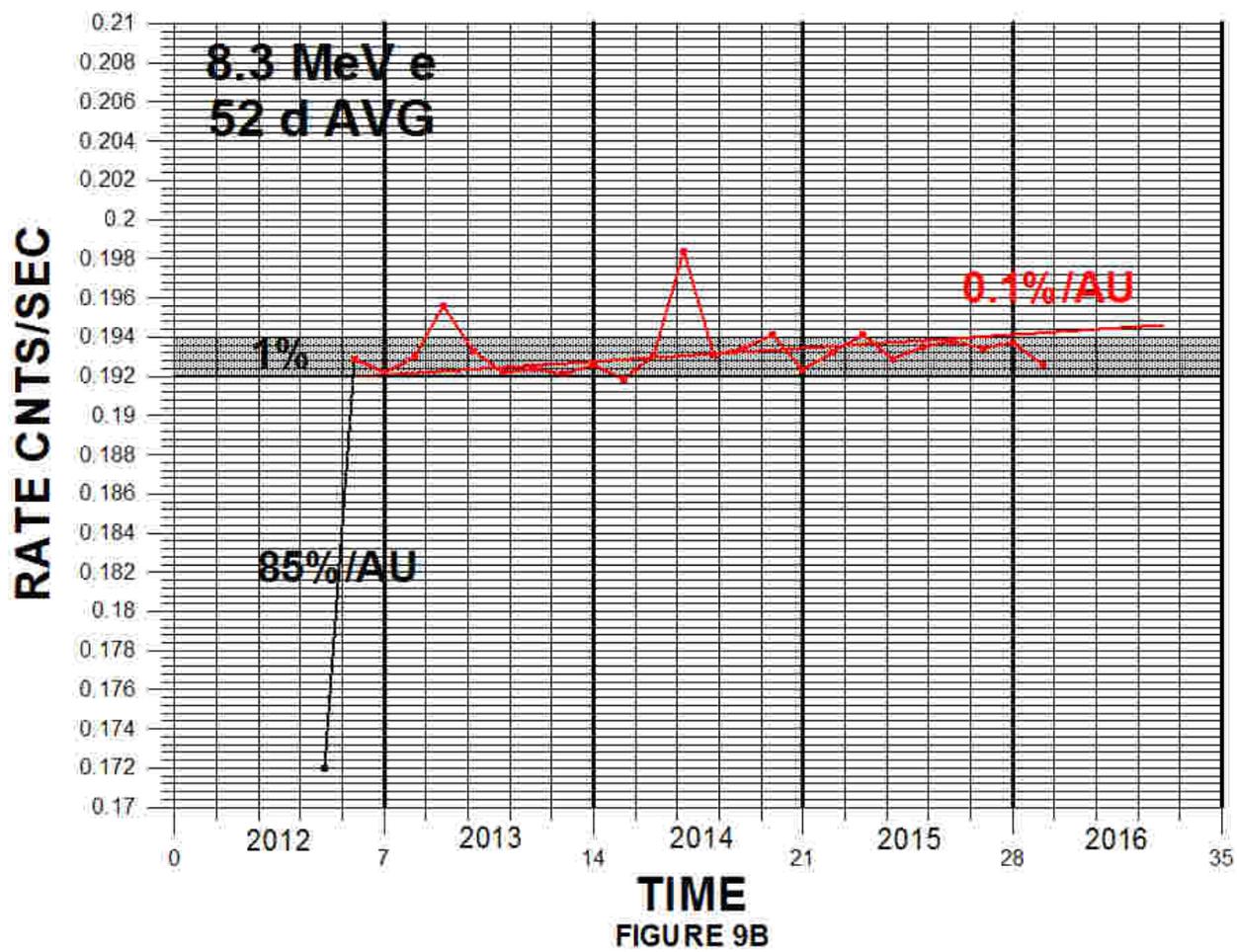

FIGURE 9B

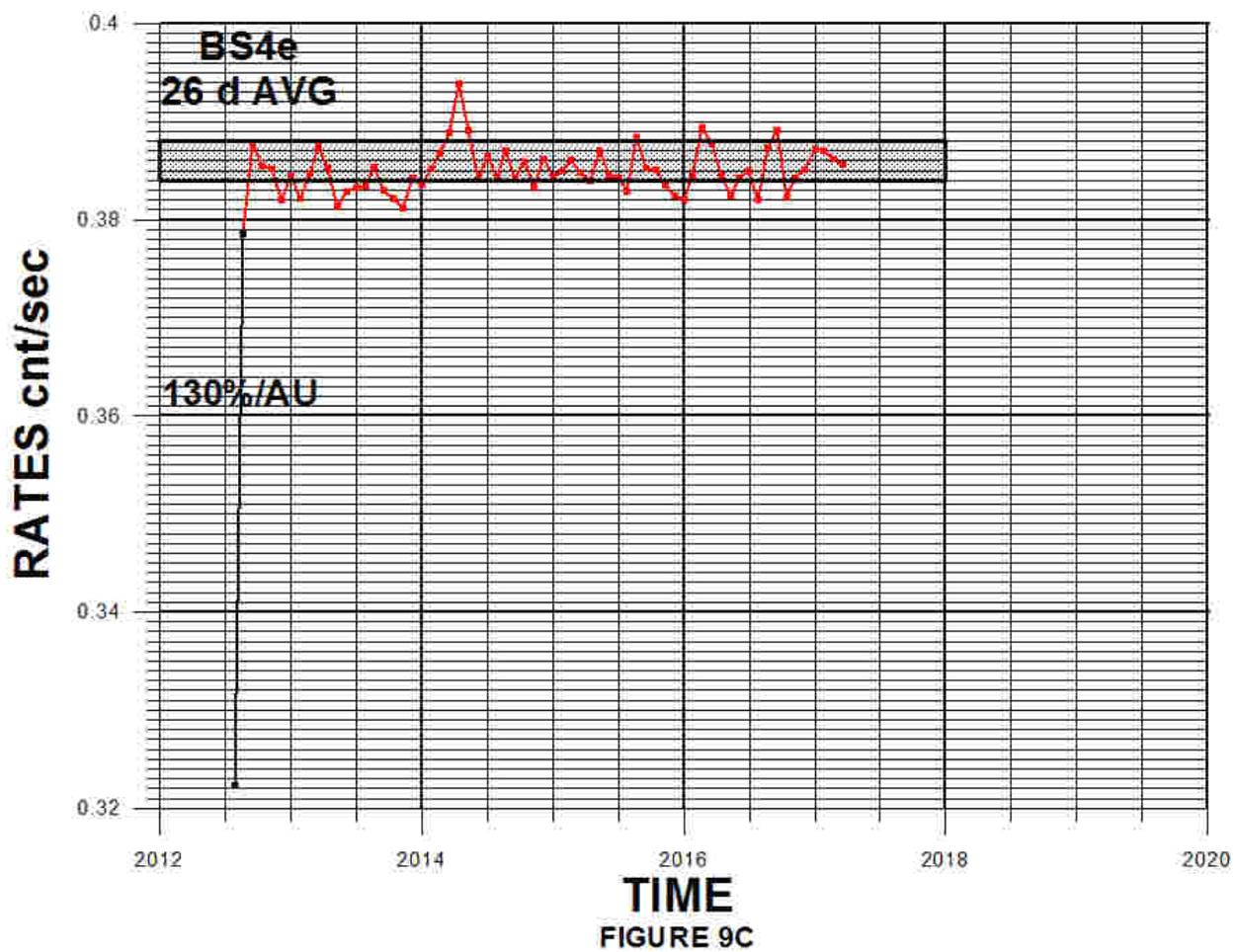

**FIGURE 9C**

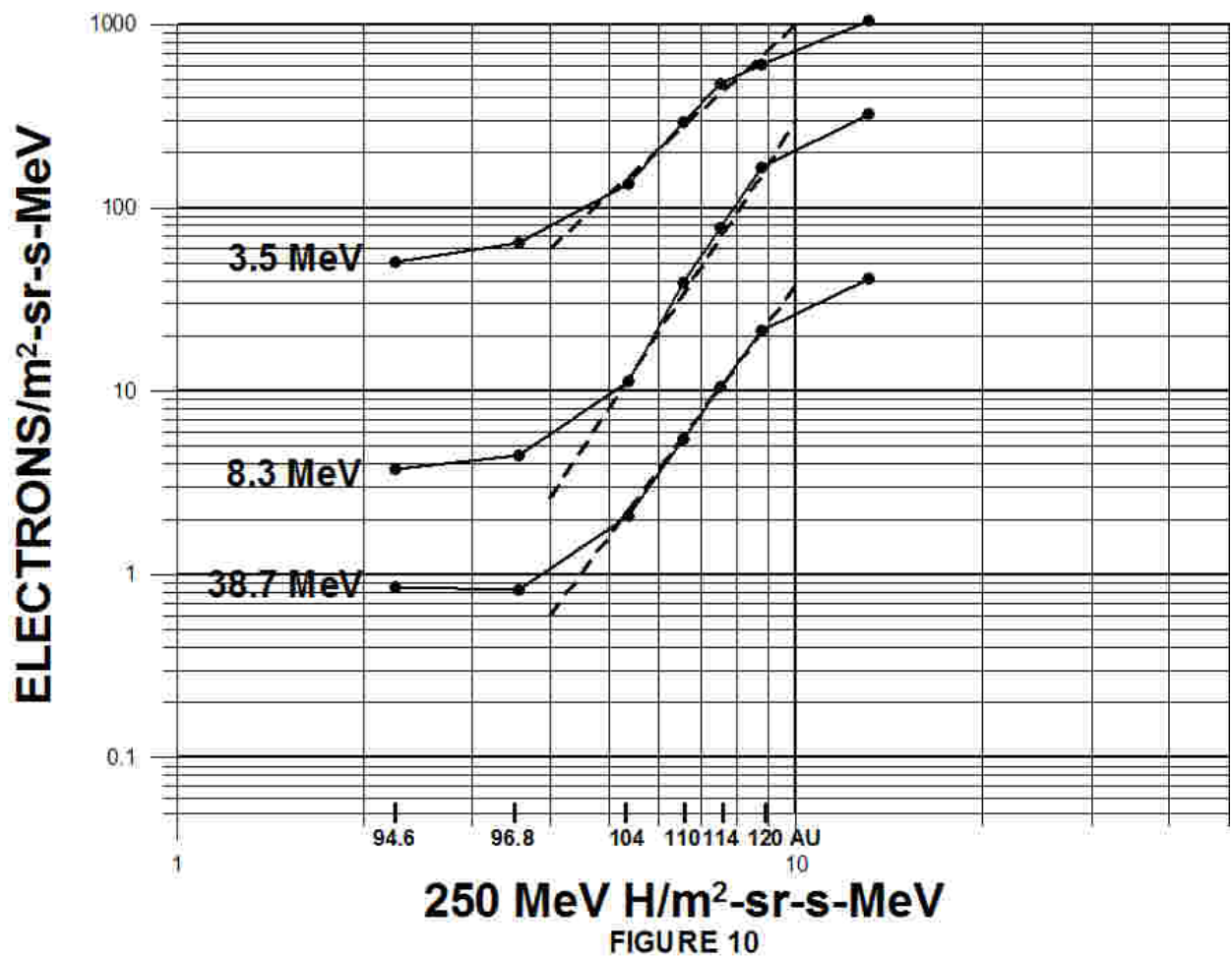

**FIGURE 10**